\documentclass[a4paper,fleqn]{cas-dc}

\usepackage{lipsum}

\usepackage[numbers]{natbib}

\xspaceaddexceptions{]}
\def\tsc#1{\csdef{#1}{\textsc{\lowercase{#1}}\xspace}}
\tsc{WGM}
\tsc{QE}
\tsc{EP}
\tsc{PMS}
\tsc{BEC}
\tsc{DE}

\begin{document}
\let\WriteBookmarks\relax
\def\floatpagepagefraction{1}
\def\textpagefraction{.001}
\shorttitle{}
\shortauthors{Morfouace et al.}

\title [mode = title]{Constraining the symmetry energy with heavy-ion collisions and Bayesian analyses}                      




\author[1]{P. Morfouace}[]
\ead{pierre.morfouace2@cea.fr}
\cormark[1]

\author[1]{C.Y. Tsang}[]
\author[2]{Y. Zhang}[]
\author[1]{W. G. Lynch}[]
\author[1]{M.B. Tsang}[]
\author[1]{D.D.S Coupland}[]
\author[1]{M. Youngs}[]
\author[3]{Z. Chajecki}[]
\author[3]{M.A. Famiano}[]
\author[5]{T.K. Ghosh}[]
\author[1]{G. Jhang}[]
\author[4]{Jenny Lee}[]
\author[6]{H. Liu}[]
\author[1]{A. Sanetullaev}[]
\author[1]{R. Showalter}[]
\author[1]{J. Winkelbauer}[]

\address[1]{National Superconducting Cyclotron Laboratory and Department of Physics and Astronomy, Michigan State University, East Lansing, MI 48824, USA}
\address[2]{China Institute of Atomic Energy; Beijing, 102413, PRC}
\address[3]{Department of Physics, Western Michigan University, Kalamazoo, Michigan 49008, USA}
\address[4]{Department of Physics, The University of Hong Kong, Hong Kong, China}
\address[5]{Variable Energy Cyclotron Centre, 1/AF Bidhannagar, Kolkata 700064, India}
\address[6]{Texas Advanced Computing Center, University of Texas, Austin, Texas 78758, USA}


\credit{Data curation, Writing - Original draft preparation}

\cortext[cor1]{Corresponding author}

\begin{abstract}[S U M M A R Y]
Efficiency corrected single ratios of neutron and proton spectra in central $^{112}$Sn+$^{112}$Sn and $^{124}$Sn+$^{124}$Sn collisions at 120 MeV/u are combined with double ratios to provide constraints on the density and momentum dependencies of the isovector mean-field potential. Bayesian analyses of these data reveal that the isoscalar and isovector nucleon effective masses, $m_s^* - m_v^*$ are strongly correlated. The linear correlation observed in  $m_s^* - m_v^*$ yields a nearly independent constraint on the effective mass splitting $\Delta m_{np}^*= (m_n^*-m_p^*)/m_N = -0.05_{-0.09}^{+0.09}\delta$. The correlated constraint on the standard symmetry energy, $S_0$ and the slope, $L$ at saturation density yields the values of symmetry energy $S(\rho_s)=16.8_{-1.2}^{+1.2}$ MeV at a sensitive density of $\rho_s/\rho_0 = 0.43_{-0.05}^{+0.05}$. 
\end{abstract}


\maketitle

Connecting the properties of matter within neutron stars to the properties of nuclei on earth presents both opportunities and challenges. The ability to study the symmetry energy by nuclear measurements in the laboratory presents a definite opportunity. On the other hand, the large difference between the asymmetry of matter within nuclei and that of neutron stars presents a definite challenge. In nuclei, the Coulomb forces shift the energy minimum to more neutron-rich isotopes in heavy nuclei, but the symmetry energy shifts the energy minimum to more symmetric isotopes. Consequently, the interplay of Coulomb and symmetry energies limit the neutron number $N$ available for elements of proton number $Z$ to a narrow range about $N$ and the asymmetries $\delta=(N-Z)/(N+Z)$ of nuclei remain less than 0.25. However, inside neutron stars, the neutron fraction can reach above 90\% at normal density under $\beta$ equilibrium conditions. This vastly increases the importance of probing the symmetry energy and understanding its effects in the laboratory over wide range of densities and asymmetries.

\begin{figure*}
\includegraphics[width=\textwidth]{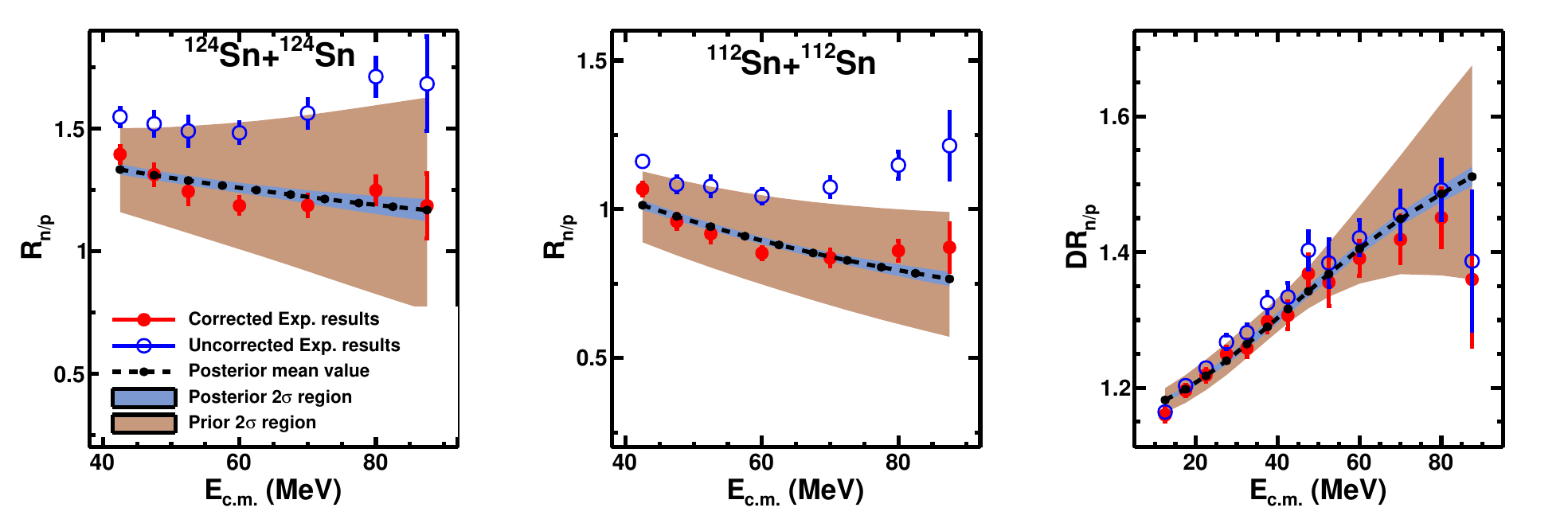}
\caption{\label{fig:Rnp} Single neutron over proton ratio for $^{124}$Sn+$^{124}$Sn and $^{112}$Sn+$^{112}$Sn at 120 MeV/u. The open blue circles show the ratio before correction and the full red points show the ratio after correction due to multiple scattering and nuclear reaction losses. The brown shaded area corresponds to the prior $2\sigma$ region of the 49 sets of ImQMD calculation spanning the parameters space while the blue colored area correspond to the posterior $2\sigma$ region.}
\end{figure*}

Recent observation of a neutron star-merger event \cite{abb17} yields the first glimpse of neutron-star properties such as tidal deformability that are governed by the nuclear equation of state (EoS). The EoS at zero temperature is the sum of the symmetry energy and the energy for symmetric matter with equal neutron and proton density, $\rho_n=\rho_p$. For neutron stars, the density dependence of the symmetry energy, $E_{sym}=S(\rho)\delta^2$, strongly influences the relationship between pressure and the density, $\rho=\rho_n+\rho_p$, of stellar matter and thus, the neutron star mass-radius relationship \cite{lat04, ste05} as well as the nuclear lattice and nucleonic gas within the inner crust, the boundary between the core and the inner crust and the nature of lattice or pasta structures of nuclei. Laboratory constraints have been obtained on the EoS \cite{dan02} and on the momentum dependence of the mean-field potentials for symmetric matter \cite{li13, li15}. Present efforts to constrain $S(\rho)$ have focused on the first two coefficients $S_0$ and $L$ in the Taylor expansion of $S(\rho)$ around the saturation density $\rho_0$,
\begin{equation}
S(\rho) = S_0+\frac{L}{3\rho_0}(\rho-\rho_0)+O((\rho-\rho_0)^2)
\label{eq:Srho}
\end{equation}
Information about $S_0$ and $L$ have been obtained from analyses of the masses \cite{kor10, bro13}, Pygmy Dipole Resonances (PDR) \cite{kli07,wie11,car10}, Electric dipole polarizability \cite{ros13, tam11}, neutron skin thickness \cite{zha13}, Isobaric Analog States (IAS) \cite{dan17} and isospin diffusion in heavy-ion collisions \cite{tsa09}. These analyses result in positively-correlated constraints on $S_0$ and $L$. Depending on the experimental condition, the slope of the correlation between $S_0$ and $L$ are different. That is because the slope is a signature of the sensitive density being probed by a given laboratory experiment \cite{lyn17}.

In addition to the density dependence of the symmetry potential, the nuclear mean-field potential has momentum dependencies from the Fock exchange term, finite range and correlation effects \cite{li04, liu02, bru55, mah85, far01, zuo02, hof01, gre03, van05}. The neutron and proton effective masses associated with these effects influence many of the thermal properties of hot proto-neutron stars formed in core-collapse supernovae \cite{lat04, ste05, jan12, bet90}. The mean-field potential contains an isoscalar effective mass $m_s^*$ that is reduced in nuclei from the nucleon mass $m_N$ by approximately $\frac{m_s^*}{m_N}\approx0.65-0.75$ \cite{li04, bru55, mah85}. Furthermore, momentum dependencies in the isovector (symmetry) mean-field potential will cause the neutron and proton effective masses to differ \cite{liu02}. This effect strongly modifies the cooling of neutron stars via neutrino emission \cite{bal14}.

\begin{table}[pos=h]
\begin{tabular}{|c|}        
 Parameter range  			 \\ \hline \hline
$25.7 \le S_0 \le 36$ (MeV) 	\\ 
$32 \le L \le 120$ (MeV)		\\ 
$0.6 \le m_s^*/m_N \le 1.0$	\\ 
$0.6 \le m_v^*/m_N \le 1.2$	\\ 
\end{tabular}
\caption{\label{tab:ImQMD_param} Model parameter values for prior distribution. 49 sets of calculation have been performed within this 4D model space using a Latin hyper-cube sampling.}
\end{table}

Theoretical calculations and commonly used effective interactions differ regarding the sign and magnitude of the effective-mass splitting $\Delta m_{np}^*=\frac{m_n^*-m_p^*}{m_N}$. Positive values for the mass splitting are expected from Landau Fermi liquid theory \cite{sjo76} and this sign appears to be consistent with recent fits to the energy dependence of the nucleon elastic scattering that obtain $\Delta m_{np}^*=(0.27\pm0.25)\delta$ \cite{li13}. Calculations predict the effective-mass splitting to increase strongly with density, an effect that becomes increasingly important in astrophysical environments such as neutron stars and in central heavy-ion collisions \cite{zha14,riz05,tor10}.

\begin{figure*}
\begin{center}
\includegraphics[width=\textwidth]{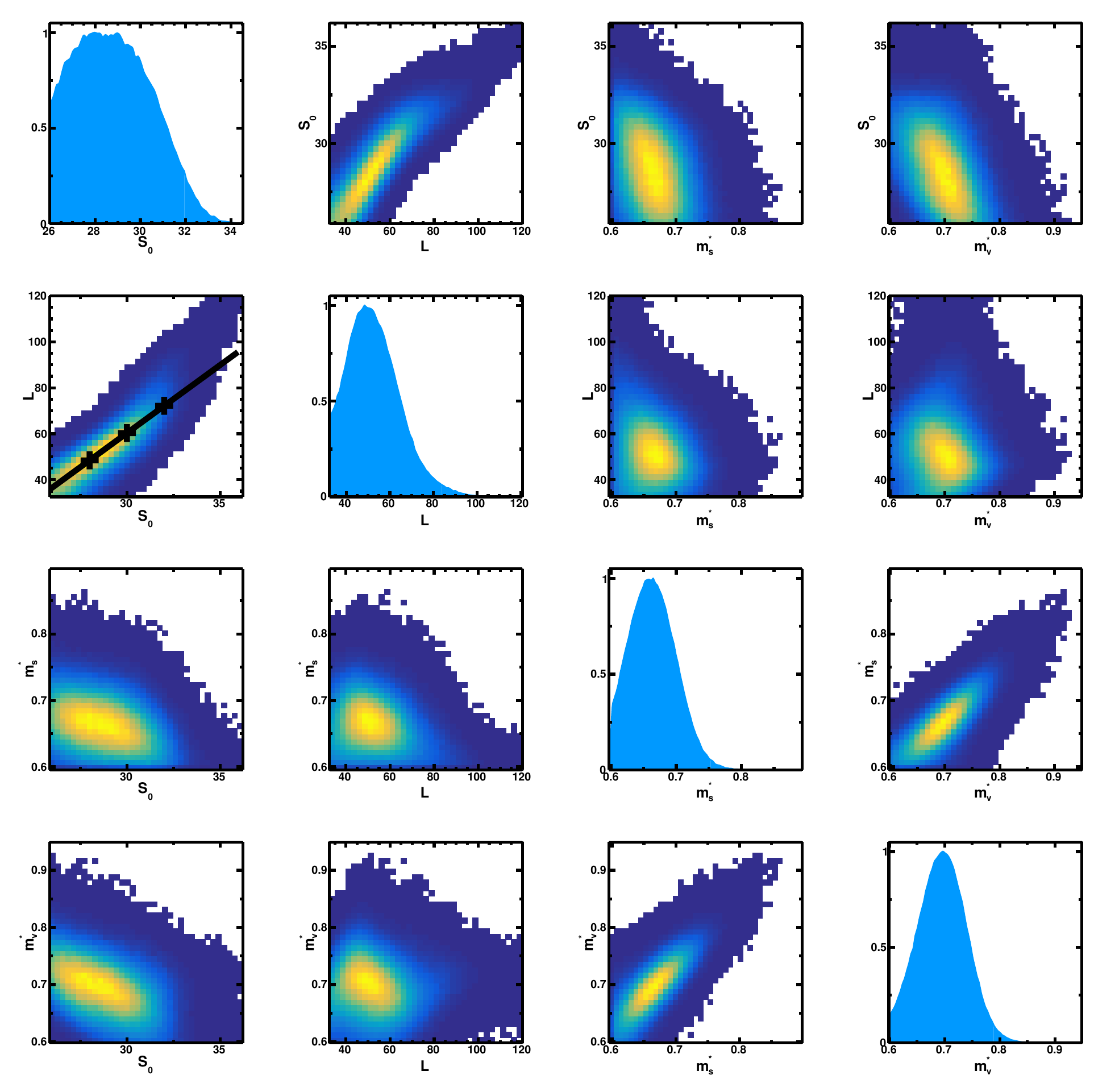}
\caption{\label{fig:MADAI} The posterior likelihood for two parameters showing the constrains on those parameters. The projections of those plots correspond to the one-dimensional spectrum illustrating how a given parameter is constrained by the data.}
\end{center}
\end{figure*}

\begin{figure*}
\begin{center}
\includegraphics[width=\textwidth]{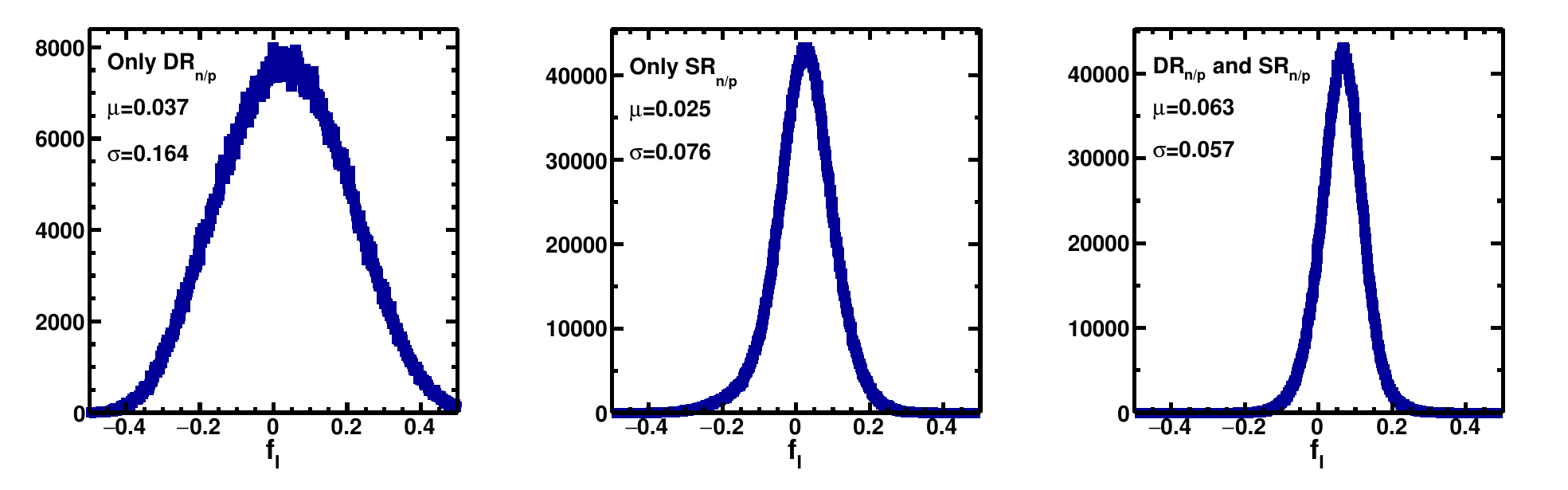}
\caption{\label{fig:fi} Prior of the $f_I$ distribution when using only the double ratio $DR_{n/p}$ in the Bayesian analysis (left), only the single ratio $R_{n/p}$ (center) and the double and single ratio combined (right).}
\end{center}
\end{figure*}

In this paper, we probe both the density and the momentum dependence of the symmetry energy via measurements of neutron and light charged particle spectra in central $^{124}$Sn+$^{124}$Sn  and $^{112}$Sn+$^{112}$Sn collisions at E/A=120 MeV. Details about this experiment can be found in ref.~\cite{cou16}. Light-charged particles were measured in the Large-Area Silicon Strip Array (LASSA) \cite{dav01} placed 20 cm from the target covering the polar angle range of $23^{\circ} < \theta_{lab} < 57^{\circ}$ with a 0.9$^{\circ}$ angular resolution. Neutrons were measured by the two walls of the MSU Large-Area Neutron Array (LANA) \cite{zec97} at 5 and 6 m from the reaction target. The LANA spanned polar angles of $15^{\circ} < \theta_{lab} < 58^{\circ}$ with an angular resolution of $0.8^{\circ}$ to $0.9^{\circ}$. Neutrons were distinguished from $\gamma$ rays by pulse shape discrimination and from charged particles by use of a charged-particle veto array of BC-408 plastic scintillator detectors placed between the target and the neutron walls. In order to avoid systematic uncertainties associated with the single ratios, ref.~\cite{cou16} reported only spectral double ratios described below. In this work, we carefully correct the charged-particle spectra for reaction and scattering losses in the CsI(Tl) crystals causing a mis-identification of charged particles in the LASSA with an estimated accuracy of $\pm 2\%$ \cite{mor17}, while the neutron efficiency has been carefully estimated in Ref.~\cite{cou16} by using the SCINFULQMD Monte Carlo code \cite{sat06}. In general, most transport models have difficulty reproducing the relative abundances of light isotopes produced as the system expands and disassembles. Following ref.~\cite{cou16} we calculated the coalescence invariant (CI) neutron and proton spectra by combining the free nucleons with those bound in light isotopes with $1<A<5$. 

The open blue circles in the left and middle panel of Fig.~\ref{fig:Rnp} show the uncorrected ratios of the coalescence-invariant neutron spectra divided by the coalescence-invariant proton spectra for the $^{124}$Sn+$^{124}$Sn and $^{112}$Sn+$^{112}$Sn. The solid red points show the corresponding efficiency-corrected single ratios. Both spectra have been transformed to the center of mass as described in ref.~\cite{cou16}. The right panel shows the double ratios obtained by taking the single ratios in the left panel and dividing by the corresponding ratios in the middle panel. The efficiency corrected double ratios are consistently lower than the uncorrected double ratios but still within the experimental uncertainties. The residual differences in the double ratios come from different admixtures of the various isotopes, each with its own detection efficiency in each reaction.

The availability of the new single ratio data allows multi-parameter analysis to extract both the density and momentum dependencies of the mean-field potentials \cite{zha14, zha15}. We perform these evaluations with the ImQMD transport model of ref.~\cite{zha14}, which parameterizes the mean fields in terms of standard Skyrme parameterizations. We focus on four quantities: $S_0$ and $L$, which describe the density dependence of the mean-field potential, the isoscalar effective mass $m_s^*$ and the isovector effective mass $m_v^*$. Using a Bayesian Markov Chain Monte Carlo statistical analysis software, we explored the four-dimensional parameter space as listed in Tab.~\ref{tab:ImQMD_param}. Other parameters in the Skyrme interactions, the in-medium nucleon-nucleon cross section and Pauli blocking algorithm were kept at the default values given in Ref.~\cite{zha14, zha05, zha06, zha07}.

All calculations were performed at impact parameter $b=2$ fm, corresponding to central collisions. That choice is justified because the calculated values for the single ratio $R_{n/p}$ are relatively insensitive to impact parameter, changing negligibly ($<3$\%) over the range $b=2-6$ fm. For each system ($^{124}$Sn+$^{124}$Sn and $^{112}$Sn+$^{112}$Sn at 120 MeV/u), 49 parameter sets have been selected on a Latin hyper-cube to span the parameter space listed in Tab.~\ref{tab:ImQMD_param}. The $i^{th}$ set of parameter values in our parameter space can be represented by a 4D vector $\vec{x_i} = \{S_0, L, m_s^*, m_v^* \}$. For each of these 49 sets we run the full ImQMD model and the results of those calculation will serve to train the emulator that models the ImQMD calculations \cite{nov14}. Partly due to the steep decrease in high energy particle, our empirical studies indicate that a minimum of 200000 events per set is needed to stabilize the current results.
 
From Bayes theorem the probability $\mathcal{P}_{post}(\vec{x_i},y_{exp})$, for theoretical values $\vec{x_i}$ to be correct is given by $\mathcal{P}_{post}(\vec{x_i},y_{exp}) \propto \mathcal{P}_{likelihood}(y_{exp},\vec{x_i})\mathcal{P}_{prior}(\vec{x_i})$, where $\mathcal{P}_{prior}$ is the assumed probability distribution of the parameter set $\vec{x_i}$ determined from other information prior to comparing to the experiment. We take the conditional probability of getting a measured set of data given $\vec{x_i}$ to be the likelihood function $\mathcal{L}(\vec{x_i})$:
\begin{equation}
\mathcal{L}(\vec{x_i}) \approx exp\Big( -\sum_a \frac{(y_a^M(\vec{x_i}) - y_a^{exp} )^2}{2\sigma_a^2} \Big).
\end{equation}
In this approach, we compare the model value $y_a^M(\vec{x_i})$ for the experimental measurement $y_a^{exp}$, and the uncertainties incorporate both the experimental and the theoretical ones. 

The prior probability distributions for $S_0$, $L$ and $m_v^*$ are assumed to be uniform within the model space. For $m_s^*$ which has been shown to have a value close to 0.7, we assume a Gaussian distribution centered as 0.7 with a width of 0.05. We then use the efficiency corrected experimental data shown in Fig.~\ref{fig:Rnp} to evaluate the post probability distribution using Markov Chain Monte Carlo sampling (MCMC) which is implemented in the PyMC library \cite{pymc}. The AutoGrad package \cite{autograd} is used in order to train the emulator by minimizing the emulated error. The brown shaded area in the three panels of Fig.~\ref{fig:Rnp} show the extreme values for the 49 sets of the ImQMD calculation corresponding to the prior distribution of the theoretical parameters, while the blue area show the posterior $2\sigma$ region for the four fitted parameters. The diagonal panels in Fig.~\ref{fig:MADAI} show the posterior distribution for the four parameters. Some of the parameters are highly correlated. For example, the two-dimensional plot in the upper left panels show that there is a strong correlation between $S_0$ and $L$, that has been observed in previous studies \cite{hor14}. There is also a strong correlation between $m_s^*$ and $m_v^*$ with $m_s^*/m_N=0.67\pm 0.03$ and $m_v^*/m_N=0.69\pm 0.04$. Using the following relationship for the effective mass splitting $\Delta m_{np}^*$
\begin{equation}
\begin{array}{c}
f_I=\Big(\frac{m_N}{m_s^*} - \frac{m_N}{m_v^*}\Big)=  \frac{1}{2\delta}\Big(\frac{m_N}{m_n^*} - \frac{m_N}{m_p^*}\Big)\\
f_I \approx - \frac{1}{2\delta} \Delta m_{np}^*\Big( \frac{m_N}{m_s^*} \Big)^2,
\end{array}
\end{equation}
one gets $f_I=0.06\pm 0.06$ or $\Delta m_{np}^*=(-0.05\pm 0.09)\delta$. If the single ratio data are removed from the Bayesian analysis, only the correlation between $m_s^*$ and $m_v^*$ remains and the $f_I$ distribution becomes broader but still centers closer to 0 ($\mu=0.037$ and $\sigma=0.164$) as shown in the left panel of Fig.~\ref{fig:fi}. Within the uncertainties, this is consistent with the double ratio analysis \cite{cou16}. Inclusion of the single ratio data allows the extraction of the $S_0$ and $L$ correlation and the tightening of the effective mass splitting constraint for $f_I$ ($\mu=0.063$ and $\sigma=0.057$) as shown in the right panel of Fig.~\ref{fig:fi}. This uncertainty reflects the relatively small influence of the effective mass splitting on the single and double ratios for nucleon energies less than $E_{c.m.}=100$ MeV. We note that a much larger sensitivity is expected for nucleons emitted at higher incident energies; measurements such higher energies should lead to more definite result \cite{zha14,zha15,riz05}. The current results are consistent with the lower bound on $\Delta m_{np}^*$ obtained from elastic scattering in ref.~\cite{li15}, but lower than the result obtained by a statistical analysis of values published for $S_0$ and $L$ in ref.~\cite{li13}. As discuss below, such $S_0$ and $L$ values require model dependent extrapolations of $S_0$ and $L$ from measurements that are sensitive for $S(\rho)$ at much lower densities. It is unlikely that the theoretical uncertainties of those extrapolations are fully reflected in the error bars of ref.~\cite{li15}.

\begin{figure}
\begin{center}
\includegraphics[scale=0.4]{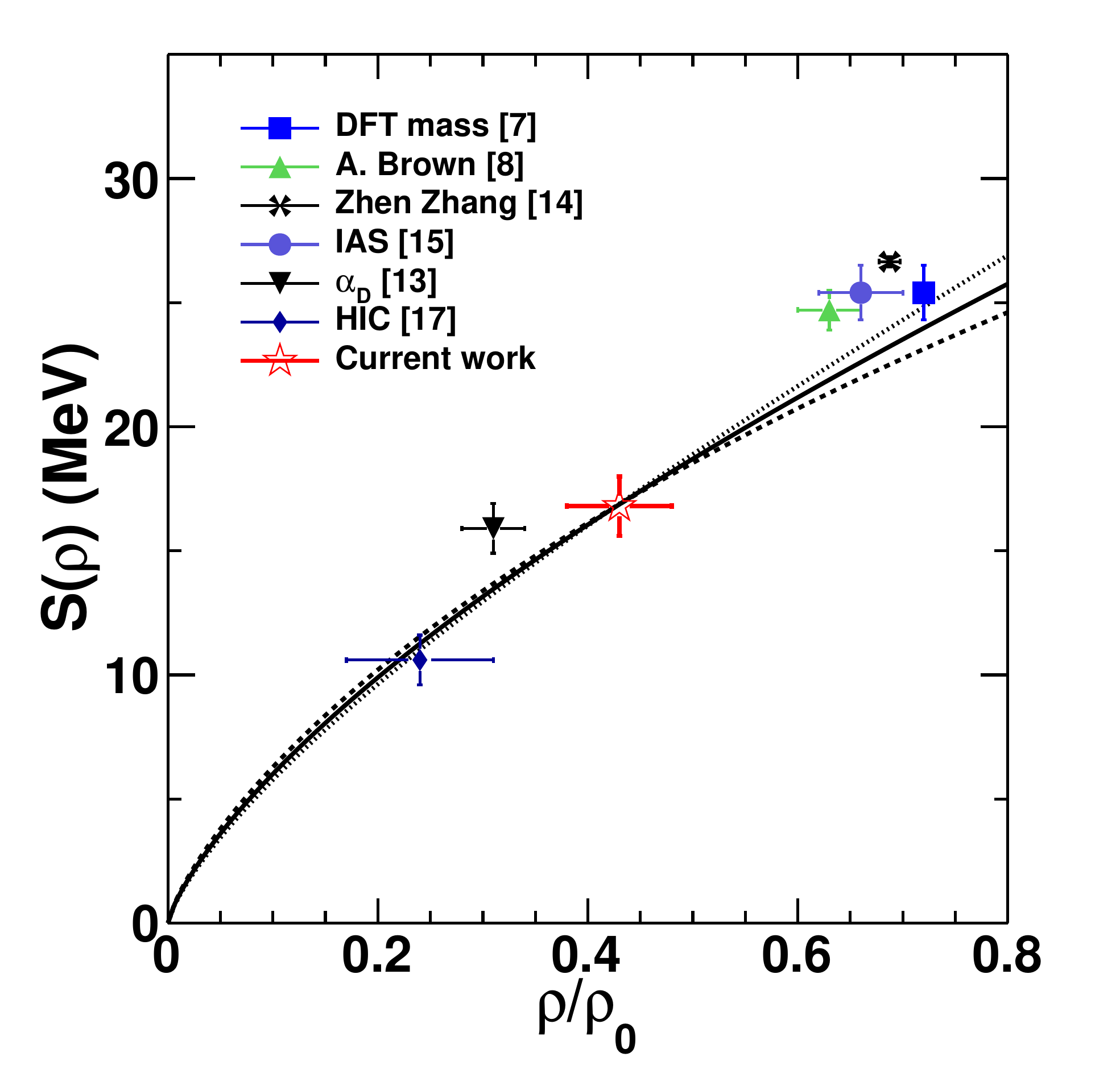}
\caption{\label{fig:density} The three black lines correspond to the symmetry energy $S(\rho)$ versus the density for different values of $S_0$ and $L$ following the slope  shown by the black crosses in Fig.~\ref{fig:MADAI}. The open red star corresponds to the cross-over point of the black lines shown in the left plot corresponding to the sensitive density $\rho_s/\rho_0 = 0.43_{-0.05}^{+0.05}$ with $S(\rho_s)=16.8\pm1.2$ MeV.}
\end{center}
\end{figure}

 
We use the overlap method described in \cite{lyn17} to determine the sensitive density and symmetry energy from our analysis. First we choose three points (black crosses in Figure~\ref{fig:MADAI}) along the best-fit linear correlation between $S_0$ and $L$. The three black curves plotted in Fig.~\ref{fig:density} represent $S(\rho)$ calculated as a function of density. $S(\rho)$ corresponds to the homogenous hadron EoS that require the symmetry energy to be zero at zero density. The value of the parameters to calculate those three curves are listed in Table.~\ref{tab:par_value}. The three black curves cross over at $\rho_s/\rho_0=0.43_{-0.05}^{+0.05}$ with $S(\rho_s)=16.8_{-1.2}^{+1.2}$ MeV which is plotted as the open red star in Fig.~\ref{fig:density}. This may indicates that even though we can obtain the density dependence of the symmetry energy over a very large range of density regions using the Bayesian analysis, the region best explored by the experiment is limited. The other symbols are the extracted symmetry energy obtained in ref.\cite{lyn17} from masses \cite{kor10, bro13} and isobaric analog sates \cite{dan17} around $0.67\rho_0$. The dipole polarizability measurements \cite{tam11} and the isospin diffusion data sit at $0.32\rho_0$ and $0.25\rho_0$, respectively. 

To exploit the full potential of the Bayesian analysis to extract multi-parameters in the Equation of state, both the data and the theoretical model need to be improved. The Bayesian analysis depends highly on the theoretical model used. Specifically, the current ImQMD model and most transport models have trouble reproducing the shape of the single ratio especially for low energy particles less than 40 MeV/u. There is also a discrepancies in the single ratios at high nucleon energy. Data with better quality and extended to high energy explore the region with higher sensitivity to the effective mass \cite{cou16}. While this analysis considers uncertainties in the predictions for the double and single ratios due to uncertainties in values for $S_0$, $L$, $m_s^*$ and $m_v^*$, other model uncertainties in the functional form for the effective mass terms \cite{xu16, zha18} are more difficult to define and therefore are not explored in this paper. We note that if any effect leads to a renormalization of the calculations of the order of 5\%, the correlation between $m_v^*$ and $m_s^*$ is significantly altered while the correlation between $S_0$ and $L$ and their associated constraints remain. This is partly because the effects of effective mass splitting are much smaller than the effect of the density dependence of the symmetry energy. Obtaining accurate data at higher nucleon energies are the main goals of a series of recent experiments \cite{nscl_exp}.

\begin{table}[pos=h]
\begin{tabular}{c | c | c | c | c}        
$S_0$ (MeV)	& $L$ (MeV)	& $m_s^*/m_N$ & $m_v^*/m_N$	& $f_I$ \\ \hline \hline
28		& 48.0	& 0.67  & 0.72			& 0.098\\
30		& 61.8	& 0.65  & 0.69			& 0.080\\
32		& 75.6	& 0.63  & 0.66			& 0.076\\
\end{tabular}
\caption{\label{tab:par_value} Parameter value used to calculate the three symmetry energy represented by the three black curves in Fig.~\ref{fig:density}, that correspond to the three black crosses in Fig.~\ref{fig:MADAI}.}
\end{table}

In summary, we have presented new results for the single ratios of coalescence invariant neutron/proton spectra from central $^{124}$Sn+$^{124}$Sn and $^{112}$Sn+$^{112}$Sn collisions at 120 MeV/u. We have shown that the Bayesian analyses can be used for multivariable analysis. However, the results from the analysis is model dependent. Nonetheless, the analysis show a strong correlation between the values for $S_0$ and $L$, which is absent if the single ratio data is not included in the analysis. Together with the double ratio, these data also provide significant constraints on the effective-mass splitting around half saturation density which is near the crust-core transition density in neutron star \cite{lat16}. This region also serves as a bridge between the density regions investigated with nuclear structure experiments ($\approx 0.7 \rho_0$) and very-low-energy heavy-ion collisions ($\le 0.3 \rho_0$) which is important for the question of clustering at very low density that corresponds to density relevant for the neutrino sphere physics.

We thank Scott Pratt, Earl Lawrence and Michael Grosskopf for many helpful discussions on the Bayesian analysis. This research is supported by the National Science Foundation under Grant No. PHY-1565546. We acknowledge the computational resources provided by the Austin Advance Computer Center and the Institute for Cyber-Enabled Research at Michigan State University.





\makeatletter

\def\pct{\expandafter\@gobble\string\%}

\immediate\write\@auxout{\pct\space This is a test line.\pct }

\end{document}